# Silicon-Compatible Ionic Control over Multi-State Magnetoelectric Phase Transformations in Correlated Oxide System

Xuanchi Zhou [1, 2] †*, Jiahui Ji [1] †, Wentian Lu [1, 2] †, Huihui Ji [1, 2], Chunwei Yao [1], Xiaohui Yao [1], Xiaomei Qiao [1], Guowei Zhou [1, 2] *, Xiaohong Xu [1, 2] *

[1] *Key Laboratory of Magnetic Molecules and Magnetic Information Materials of Ministry of Education & School of Materials Science and Engineering, Shanxi Normal University, Taiyuan, 030031, China*
[2] *Research Institute of Materials Science, Shanxi Key Laboratory of Advanced Magnetic Materials and Devices, Shanxi Normal University, Taiyuan 030031, China*

*Authors to whom correspondence should be addressed: xuanchizhou@sxnu.edu.cn (X. Zhou), zhougw@sxnu.edu.cn (G. Zhou), and xuxh@sxnu.edu.cn (X. Xu).

† X. Zhou, J. Ji, and W. Lu contributed equally to this work.

**Abstract**

Realizing room-temperature ferromagnetic insulators, critical enablers for low-power spintronics, is fundamentally challenged by the long-standing trade-off between ferromagnetic ordering and indirect exchange interactions in insulators. Ionic evolution offers tempting opportunities for accessing exotic magnetoelectric states and physical functionality beyond conventional doping paradigm via tailoring the charge-lattice-orbital-spin interactions. Here, we showcase the precise magneto-ionic control over magnetoelectric states in LSMO system, delicately delivering silicon-compatible weakly ferromagnetic insulator state above room temperature. Of particular note is the decoupling of ion-charge-spin interplay in correlated LSMO system, a primary obstacle in clarifying underlying physical origin, with this process concurrently giving rise to an emergent intermediate state characterized by a weakly ferromagnetic half-metallic state. Benefiting from the $SrTiO_3$ buffer layer as epitaxial template to promote interfacial heterogeneous nucleation, hydrogenation enables diverse magnetoelectric states in LSMO integrated on silicon, fully compatible with traditional semiconductor processing. Assisted by theoretical calculations and spectroscopic techniques, hydrogen-induced magnetoelectric transitions in LSMO are driven by band-filling control and suppression in double exchange interaction. Our work not only defines a novel design paradigm for exploring exotic quantum states in correlated system, with transformative potential for spintronics, but also fundamentally unveils the physical origin behind ionic evolution via disentangling the ion-charge-spin coupling.

**Key words**: Correlated oxides, Magnetoelectric coupling, Topotactic phase transformation, Hydrogenation, Spintronics;

The complex interplay of instabilities in the charge, spin, lattice and orbital degrees of freedom results in a variety of emergent physical functionality and phenomena in correlated oxide system, covering superconductivity,[1, 2] metal-insulator transition,[3] ferromagnetism [4-6] and charge or spin ordering,[7] beyond traditional semiconductors. The ability to dynamically manipulate the functional properties for correlated oxides provides tempting opportunities to access exotic electronic or magnetic states beyond traditional doping paradigm, leveraging their exceptional sensitivity to external stimuli.[8] Ionic evolution introduces an additional ion degree of freedom to adjust the electron correlations and the coupling between multiple degrees of freedom, establishing a powerful tuning knob for exploring emergent physical functionality that extensively expands the structural, electronic and magnetic phase diagram.[9] The incorporation of hydrogens, the smallest and lightest ion, triggers novel structural phase transformations in correlated oxides through O-H bonding while concurrently modifying the crystalline symmetry and bandwidth in a more controllable and reversible fashion.[10] Hydrogen-related electron doping process upsets the stability in correlated electron ground state associated with the integer electron occupation, driving electronic state evolution through band-filling control.[11] Notable examples include $VO_2$,[12] $Re$NiO$_3$ [13] and SrRuO$_3$,[14] which exemplify hydrogen-triggered multiple electronic phase modulations via directly tailoring $d$-orbital filling and configuration. Moreover, hydrogenation enables precise control over the magnetic ground states of correlated oxides via adjusting spin configurations and magnetic exchange interactions.[15] This advance not only enriches multidisciplinary applications in magnetoelectronics,[16] energy conversions,[17] superconductivity,[18] ferroelectric,[19] and artificial intelligence,[20] but also offers a fertile ground to probe ion-lattice-electron interplay in correlated electron systems.

Perovskite lanthanum strontium manganite as a prototypical family of complex oxides showcases a rich spectrum of magnetoelectric states spanning from antiferromagnetic insulator to ferromagnetic metal, tailored through strontium doping concentration.[21-24] Among this family, La$_{0.67}$Sr$_{0.33}$MnO$_3$ (denoted as LSMO) hosts an enticing ferromagnetic half-metallic (FM-HM) ground state with a Curie temperature ($T_c$) of ~370 K, tunable by using external stimuli,[25-27] which captures considerable attention in spintronics.[28-32] Typically, hydrogenation offers a feasible pathway for precisely adjusting magnetoelectric states in LSMO system through tailoring magnetic exchange interactions and electronic orbital configurations.[26, 33] Strong ion-charge-spin coupling mediates hydrogen-related magnetoelectric transitions in correlated oxide system, yet poses a fundamental challenge to comprehensively understand dominant physical mechanism. From the perspective of spintronic devices, the translation of such the rich functionality achievable in hydrogenated correlated oxides into practical device applications is fundamentally constrained by the integration difficulty with conventional silicon technology. Additionally, the critical challenge in stabilizing room-temperature ferromagnetic insulator (FM-I), prime candidates for energy-efficient spintronics leveraging both the charge and spin degrees of freedom of electrons, persists in the inherent competition between ferromagnetic ordering and indirect exchange interactions in insulator.[34] Consequently, realizing room-temperature FM-I states compatible with conventional silicon-based semiconductor processing is rather crucial for low-power spintronic devices.

In this work, we demonstrate the precise magneto-ionic control over the magnetoelectric states in correlated LSMO system that significantly expands magnetoelectric phase diagram, among which emergent room-temperature weakly ferromagnetic insulator (WFM-I) state offers new prospects for low-power spintronics. On the basis of experimental findings and theoretical calculations, hydrogen-triggered suppression in the $Mn^{3+}$-$Mn^{4+}$ double exchange interaction drives magnetic phase transition in LSMO system, which precedes the electron localization through the complete filling in $e_g$ orbital, disentangling the intertwined ion-charge-spin interaction in this correlated system. Critically, the heterogeneous integration of hydrogen-triggered multiple magnetoelectric states within LSMO system on silicon using a $SrTiO_3$ (STO) epitaxial template surmounts the primary barrier for protonic device applications. Our findings not only unveil multiple silicon-compatible magnetoelectric phase transformations in LSMO through hydrogenation but also provide fundamentally new insights into ionic evolution in correlated system via decoupling ion-charge-spin interactions.

Conventionally, hydrogenation directly donates the electron carriers into the conduction band to elevate (reduce) the Fermi level ($E_F$) that improves (depresses) the electronic conductivity in traditional $n$ ($p$)-type semiconductors through simply altering the carrier density.[35, 36] By strong contrast, proton evolution reconfigures the electronic band structure in correlated oxides to trigger the electron localization via adjusting electron correlation strength, enabling the possibility in exploring exotic electronic states unobtainable in traditional hydrogen-related phase diagram (Figure 1a). In $La_{1-x}Sr_xMnO_3$, strontium acts as an acceptor dopant to modulate the hole carriers near the $E_F$ and magnetoelectric transport behavior, among which $La_{0.67}Sr_{0.33}MnO_3$ with a $T_c$ above room temperature and an ultrahigh spin polarization of almost 100 % has captured significant attention in spintronics (Figure 1b).[28] The perovskite crystal field splits the Mn-3$d$ orbital of LSMO into a $t_{2g}$ triplet orbital and an $e_g$ doublet orbital, while the Jahn-Teller distortion arising from the lattice deformation lifts the two-fold degeneracy associated with $e_g$ orbital (Figure 1c). Hydrogen-associated band filling in empty $e_g$ orbital is poised to drive Mott orbital reconfiguration from electron-itinerant $Mn^{3+}$ ($t_{2g}^3 e_g^1$) or $Mn^{4+}$ ($t_{2g}^3 e_g^0$) ground state of LSMO toward electron-localized $Mn^{2+}$ ($t_{2g}^3 e_g^2$) state, where the fully occupied $e_g$ orbital triggers strong Mott-Hubbard electron correlations to localize electron carriers via on-site Coulomb repulsions. Beyond that, proton evolution is also expected to disturb the ferromagnetic ordering of LSMO through suppressing the $Mn^{3+}$-$Mn^{4+}$ double exchange interaction (Figure 1d). Therefore, hydrogenation provides an essential pathway to design novel electronic and magnetic states in correlated LSMO system through adjusting the ion-electron-lattice coupling, enabling exploration beyond conventional doping paradigms.

To effectively realize the proton evolution in LSMO, hydrogen spillover strategy is herein exploited, wherein the sputtered Pt as catalyst extensively reduces the energy barrier for triggering the $H_2$ (g)→$H^+$+$e^-$ soft-chemistry reaction (Figure 2a). Considering similar *a*-axis lattice constant ($a_0$) of LSMO film ($a_{0,\text{film}}$ = 3.881 Å) and STO substrate ($a_{0,\text{sub.}}$ = 3.905 Å), the minimal lattice mismatch (~0.614 %) enables the epitaxial template effect related to STO substrate, driving preferential *out-of-plane* orientation of the grown LSMO film through perovskite-to-perovskite coherent epitaxy growth. This understanding is confirmed by respective X-ray diffraction (XRD) pattern, in which the (002) plane diffraction peak associated with LSMO film just appears adjacent to the *c*-plane STO substrate (Figure 2b). Incorporating the hydrogens into the lattice of LSMO leads to the lattice expansion along *out-of-plane* direction without perturbing expected lattice framework, evidenced by the leftward shift of characteristic diffraction peak for LSMO. Such the hydrogen-triggered topotactic structural evolution is associated with the weak bonding between interstitial hydrogens and lattice oxygen that triggers the lattice expansion, an effect markedly intensified by elevating the hydrogenation temperature (Figure S1). Further consistency in structural evolution of LSMO film through hydrogenation is confirmed by comparing the reciprocal space mapping (RSM) in Figure 2c. The identical *in-plane* vector (e.g., $Q_\parallel$) between LSMO film and STO substrate verifies the coherent epitaxy growth of LSMO film deposited on the *c*-plane STO substrate. Meanwhile, an elevated *cross-plane* vector (e.g., $Q_\perp$) for the grown LSMO film relative to the STO substrate unravels the *in-plane* biaxial tensile distortion in LSMO film. With hydrogenation at 100 ºC for 1 h, the *in-plane* lattice of LSMO film is still tightly locked by the STO epitaxial template, as indicated by an identical $Q_\parallel$. By strong contrast, the reduction in the magnitude of $Q_\perp$ of LSMO through hydrogenation (e.g., from 7.758 Å$^{-1}$ to 7.736 Å$^{-1}$) demonstrates the lattice expansion of LSMO film along the *out-of-plane* direction, consistent with the above XRD results.

Hydrogen-triggered electronic phase transitions in correlated LSMO system are demonstrated by using the temperature dependences of resistivity ($\rho$-$T$ tendency) in Figure 2d, where hydrogenation progressively suppresses half-metal ground state of LSMO through enhanced electron localization via Mott-Hubbard electron correlations. As a result, hydrogenation substantially elevates the resistivity of LSMO films, with further amplification driven by increasing the hydrogenation temperature. Critically, hydrogen-induced electronic phase transition is observed for correlated LSMO under hydrogenation at 200 ºC for 1 h, driving Mott orbital reconfiguration from electron-itinerant $Mn^{3+}$/$Mn^{4+}$ ground state toward electron-localized $Mn^{2+}$ based on $t_{2g}^3 e_g^2$ state. This electronic state evolution associated with half metal-to-insulator transition through hydrogen-mediated band-filling control yields a resistivity elevation exceeding three orders of magnitude in LSMO at room temperature through electron filling in Mn-$e_g$ orbital. Such the electron localization through hydrogenation originates from the complete electron filling in $e_g$ orbital of LSMO that triggers *on-site* Coulomb repulsions that localize itinerant electrons, consistent with the need for elevated hydrogenation temperature to realize electron-localized state under sufficient hydrogen doping.

Magneto-ionic control over magnetic ground state of LSMO via hydrogenation is validated by their magnetic hysteresis loops and *M-T* tendencies shown in Figures 2e-2f. Pristine LSMO film deposited on the STO substrate exhibits a well-defined *in-plane* magnetic anisotropy, with a $T_c$ of 345 K determined from respective temperature derivative of magnetization (d*M*/d*T*) (Figure 2g). Proton evolution dramatically depresses the ferromagnetic ordering of LSMO, reducing both the saturation magnetization ($M_s$) and $T_c$, which manifests the emergence of weakly ferromagnetic (WFM) state. Notably, hydrogenated LSMO films still retain robust *in-plane* magnetic anisotropy with a preserved magnetic easy axis along the *in-plane* direction (Figure S2). Elevating the hydrogenation temperature enables more robust magnetoelectric phase control in LSMO system via enhanced proton diffusion kinetics promoted by thermal activation (Figure S3), aligning well with previous structural evolution. Despite substantial suppression of $M_s$ in LSMO hydrogenated at 300 ºC for 1 h, weakly ferromagnetic ordering still persists, evidenced by retained hysteresis loop and $T_c$ above room temperature, confirming the stabilization of a WFM state rather than AFM ordering (Figure S4). The above findings unveil unprecedented magnetoelectric coupling in correlated LSMO system via tailoring charge-lattice-spin coupling using proton evolution, wherein hydrogenation drives muti-step magnetoelectric phase modulations followed by the FM-HM→WFM-HM→WFM-I pathway.

It is in particular worthy to note that magnetic phase transition within LSMO system from FM ground state to WFM state occurs instantaneously through hydrogenation, preceding electronic phase transition that requires a thermal activation (e.g., 200 °C, 1 h). Proton evolution directly suppresses the double exchange interaction of $Mn^{3+}$-$Mn^{4+}$ pairs in LSMO to depress the ferromagnetic ordering, contrasting sharply with electronic phase modulation, where the complete filling in $e_g$ orbital via hydrogen-related electron doping just triggers the electron localization. This further reveals distinct energy landscapes governing spin and charge reconfiguration in correlated oxides via hydrogenation, decoupling the ion-charge-spin interaction. Such the asynchronous electronic and magnetic phase modulations in LSMO system through hydrogenation contributes to unravel the physical origin of magnetoelectric transitions while uncovering an emergent weakly ferromagnetic half-metallic (WFM-HM) intermediate state. Of particular note is the room-temperature WFM-I state achievable in LSMO through hydrogenation, in which electronically insulating LSMO showcases a $T_c$ exceeding room temperature, with great potential for low-power spintronics. Hydrogenation offers a feasible pathway to overcome the trade-off between ferromagnetic ordering and indirect exchange interactions in insulator via tailoring ion-charge-lattice-spin coupling, which opens up a new design paradigm in correlated oxide system.

Hydrogen-mediated magnetoelectric transitions in LSMO are highly reversible toward the pristine state via oxidative annealing or exposure to the air (Figures 3a-3b and S5). Nevertheless, residual hydrogens in deeper layers of LSMO under dehydrogenation diminish the ferromagnetic ordering and degrade half-metallic state relative to pristine counterpart. These findings reveal reversible magnetoelectric phase modulations in LSMO system via hydrogenation, switching between a FM-HM ground state and an AFM-I state. Apart from magnetoelectric transitions, hydrogen-induced structural evolution in correlated LSMO can also recover toward the pristine state through dehydrogenation, due to an ultrahigh mobility of hydrogens. To probe the hydrogen-triggered variations in chemical environment in LSMO films, X-ray photoelectron spectra (XPS) analysis was further performed, as the Mn $2p$ and O $1s$ core-level peaks shown in Figures 3c-3d, respectively. Notably, all the binding energies for LSMO films through hydrogenation obtained from XPS analysis are referenced to the standard C $1s$ peak related to the adventitious carbon that is located at 284.8 eV. The valence state of manganese can be qualitatively reflected by the binding energy of Mn $2p$ core-level peaks that are further deconvoluted into three peaks representing $Mn^{2+}$, $Mn^{3+}$ and $Mn^{4+}$ valence states, respectively.[33] Performing the hydrogenation engenders the reduction in the valence state of manganese from a mixed valence state of $Mn^{3+}$ and $Mn^{4+}$ towards $Mn^{2+}$ (Figure 3c). This observation aligns well with the previous understanding that hydrogen-related band filling triggers Mott phase transition from electron-itinerant $Mn^{3+}/Mn^{4+}$ ground state toward electron-localized $Mn^{2+}$ state based on $t_{2g}^3 e_g^2$ configuration. Further consistency is demonstrated by the O $1s$ core-level spectra in Figure 3d, where the O-H interaction (e.g., ~532.0 eV) is clearly observed for hydrogenated LSMO apart from the Mn-O interaction (e.g., ~529.9 eV), an indirect spectroscopic evidence for hydrogen incorporation within the lattice of LSMO. In particular, the effective incorporation of hydrogens is further exemplified by respective time-of-flight secondary-ion mass spectrometry (ToF-SIMS) in Figure 3e, where a marked elevation in the incorporated hydrogen concentration is identified in the LSMO film region relative to the STO substrate. Protons delivered via hydrogen spillover are poised to preferentially bond with the lattice oxygen in LSMO, forming such the weak O-H interactions, consistent with previous reports.[36, 37]

To integrate hydrogenated magnetoelectric states of LSMO on silicon, a STO buffer layer as epitaxial template was employed, where the STO buffer layer is poised to induce the interfacial heterogeneous nucleation for LSMO through structural matching (Figure 4a). This understanding is confirmed by comparing the XRD pattern for as-grown LSMO/STO/$SiO_2$/Si (001) hybrid with the one for STO/$SiO_2$/Si (001), in which case the characteristic diffraction peak representing the (002) plane of LSMO (e.g., 47.14 °) is clearly identified adjacent to the STO buffer layer. It is worthy to note that LSMO film integrated on silicon retains robust *in-plane* magnetic anisotropy, featured by a half-metallic ground state (Figure S6). With hydrogenation, topotactic structural evolution associated with the lattice expansion is analogously observed for silicon-integrated LSMO through O-H interactions, evidenced by the LSMO (002) diffraction peak shifting from right side to left side relative to the STO peak. Manipulating the ion-charge-spin coupling through hydrogenation drives magnetoelectric transitions of silicon-integrated LSMO from the FM-HM ground state to WFM-I state (Figures S7-S8), with high reproducibility (Figure S9). Notably, the intermediate WFM-HM state is absent for LSMO films integrated on silicon, attributed to an elevated material resistivity from substantially enlarged surface roughness (Figure S10). Proton evolution in LSMO/STO/SiO$_2$/Si system mirrors direct epitaxial growth on single-crystal STO substrate by inducing significant reduction of manganese valence state, with concurrent spectroscopic detection of O-H interactions. (Figure S11). Nevertheless, magneto-ionic control efficacy over magnetoelectric states in LSMO is degraded for silicon-integrated LSMO film compared to the one directly deposited on single-crystalline STO substrate (Figure S12). Silicon-compatible magneto-ionic control of magnetoelectric states in LSMO engenders novel room-temperature WFM-I phase, establishing pathways for low-power spintronic devices leveraging proton evolution.

To gain deeper insights into the physical origin behind hydrogen-mediated magnetoelectric transitions in correlated LSMO system, the spin-polarized density of states (DOS) was performed using density functional theory (DFT) calculations, as the results shown in Figure 5. It is found that an expected FM ground state is observed for pristine LSMO, aligning well with the experimental observations, as exemplified by distinct DOS for spin-down and spin-up electrons (Figure 5a). In addition, the finite DOS in the spin-up channel near the $E_F$ in LSMO contrasts sharply with the spin-down counterpart, which showcases a negligible DOS near the $E_F$, a hallmark of the half-metallic ground state (Figure S13). Introducing the hydrogens into the lattice of LSMO initially depresses the ferromagnetic ordering of LSMO, while retaining the half-metallic transport behavior (Figures 5b-5c), engendering the formation of WFM-HM state, in agreement with previously observed FM-to-WFM transitions through hydrogenation. Further elevating the incorporated hydrogen concentration to the integer number leads to the emergence of antiferromagnetic insulator (AFM-I) in hydrogenated LSMO system (Figures 5d and S14). Hydrogen-associated band filling dominantly reconfigures the Mn-$d$ orbital within LSMO system, as demonstrated by respective partial density of states (Figure S15). The magnetic transition precedes the hydrogen-induced electron localization in LSMO system, as corroborated by experimental trends, although the AFM-I state remains unattainable under current hydrogen incorporation levels. Furthermore, the direct correlation between incorporated hydrogen concentration and resulting phase modulations in LSMO via proton evolution is validated by DFT calculations (Figure S14d), exemplifying that elevating hydrogenation temperature accelerates magnetoelectric transformation via enhancing hydrogenation kinetics. Hydrogen-related multiple magnetoelectric transitions in correlated LSMO system, confirmed by both experiments and theoretical calculations, not only advance protonic device applications but also provide fundamentally new insights into Mott physics in strongly correlated systems.

In summary, hydrogen-triggered multiple magnetoelectric phase modulations are realized in correlated LSMO system via adjusting the ion-electron-lattice interplay, delicately delivering silicon-compatible WFM-I state above room temperature, with great potential in low-power spintronics. Notably, magnetic transition in LSMO system from FM to WFM state driven by the depression in $Mn^{3+}$-$Mn^{4+}$ double exchange interaction via proton evolution precedes electronic state evolution through the integer electron filling in $e_g$ orbital, as further exemplified by using DFT calculations. This advance disentangles the ion-charge-spin interplay in correlated systems, benefiting the understanding of physical origin behind hydrogen-triggered magnetoelectric transitions, while unraveling novel WFM-HM intermediate state, extending magnetoelectric phase diagram. Benefiting from the STO buffer layer, a variety of magnetoelectric states realized in LSMO system through magneto-ionic control are further integrated on silicon, compatible with conventional semiconductor processing. Our work not only showcases the precise magneto-ionic control over magnetoelectric states in correlated LSMO, yielding emergent silicon-compatible magnetoelectric states for low-power spintronics, but also provides fundamentally new insights into ionic evolution in correlated system via decoupling ion-charge-spin interplay.

**Acknowledgements**

This work was supported by the National Natural Science Foundation of China (Nos. 52401240, U24A6002, 52471203, and 12404139), Fundamental Research Program of Shanxi Province (No. 202403021212123), Scientific and Technologial Innovation Programs of Higher Education Institutions in Shanxi (No. 2024L145), and Shanxi Province Science and Technology Cooperation and Exchange Special Project (No. 202404041101030). The authors also acknowledge the beam line BL08U1A at the Shanghai Synchrotron Radiation Facility (SSRF) (https://cstr.cn/31124.02.SSRF.BL08U1A) and the beam line BL12B-b at the National Synchrotron Radiation Laboratory (NSRL) (https://cstr.cn/31131.02.HLS.XMCD.b) for the assistance of sXAS measurement.

**Figures and captions**

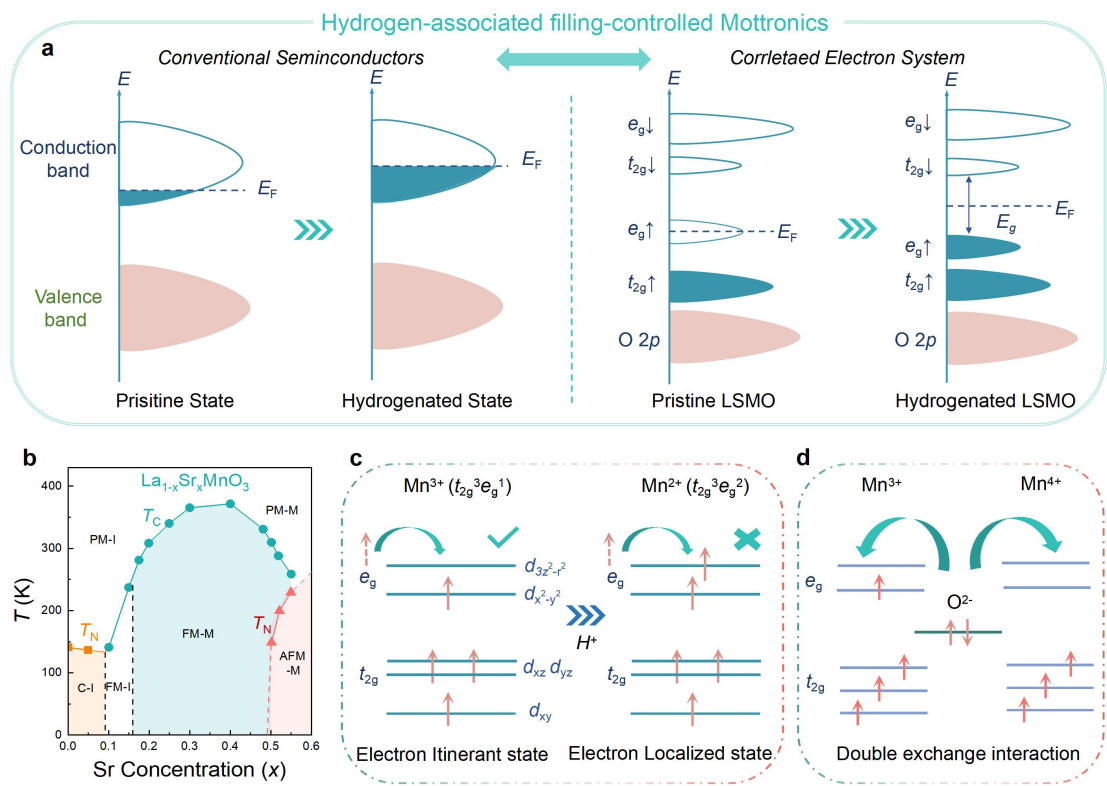

**Figure 1. Magneto-ionic control over magnetoelectric states in correlated LSMO. a**, Schematic of adjusting the electronic band structure for conventional semiconductors and correlated oxides through hydrogen-related band-filling control. **b**, Magnetoelectric phase diagram for correlated LSMO system. **c**, Schematic of orbital reconfiguration for LSMO through proton evolution. d, Schematic of hydrogen-triggered suppression in the double exchange interaction associated with $Mn^{3+}$-$Mn^{4+}$ pair.

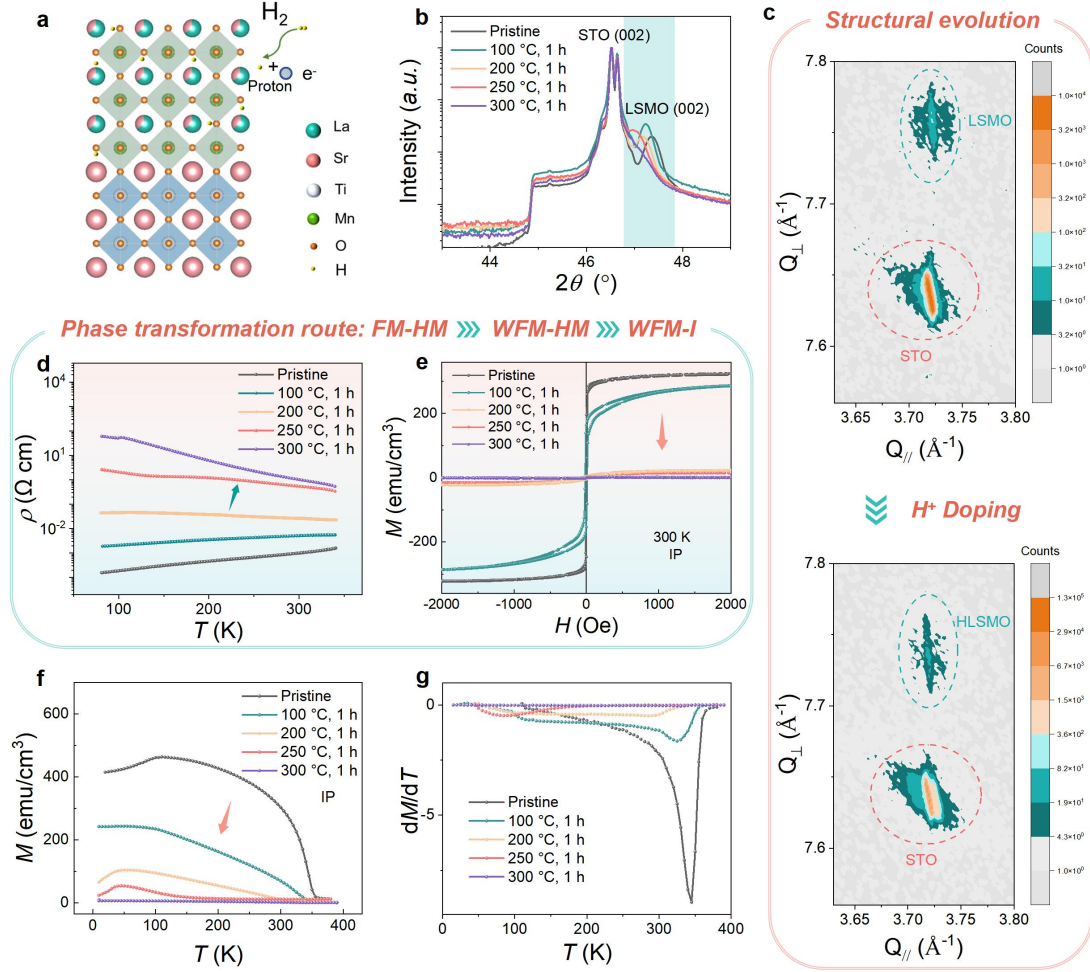

**Figure 2. Magnetoelectric phase modulations in LSMO system through proton evolution. a**, Schematic of hydrogen spillover strategy for realizing the proton evolution in LSMO. **b**, X-ray diffraction (XRD) patterns as compared for LSMO films under different hydrogenation conditions. **c**, Reciprocal space mapping (RSM) compared for LSMO/STO (001) heterostructure through hydrogenation at 100 ºC for 1 h. **d**, Temperature dependence of resistivity ($\rho$-$T$ tendency) for LSMO through hydrogenation. **e**, The *in-plane* magnetic hysteresis loops compared for pristine and hydrogenated LSMO. **f**, $M$-$T$ tendency for LSMO along an *in-plane* direction via hydrogenation via applying a magnetic field of 300 Oe. **g**, Comparing the d$M$/d$T$-$T$ tendency for LSMO film through hydrogenation.

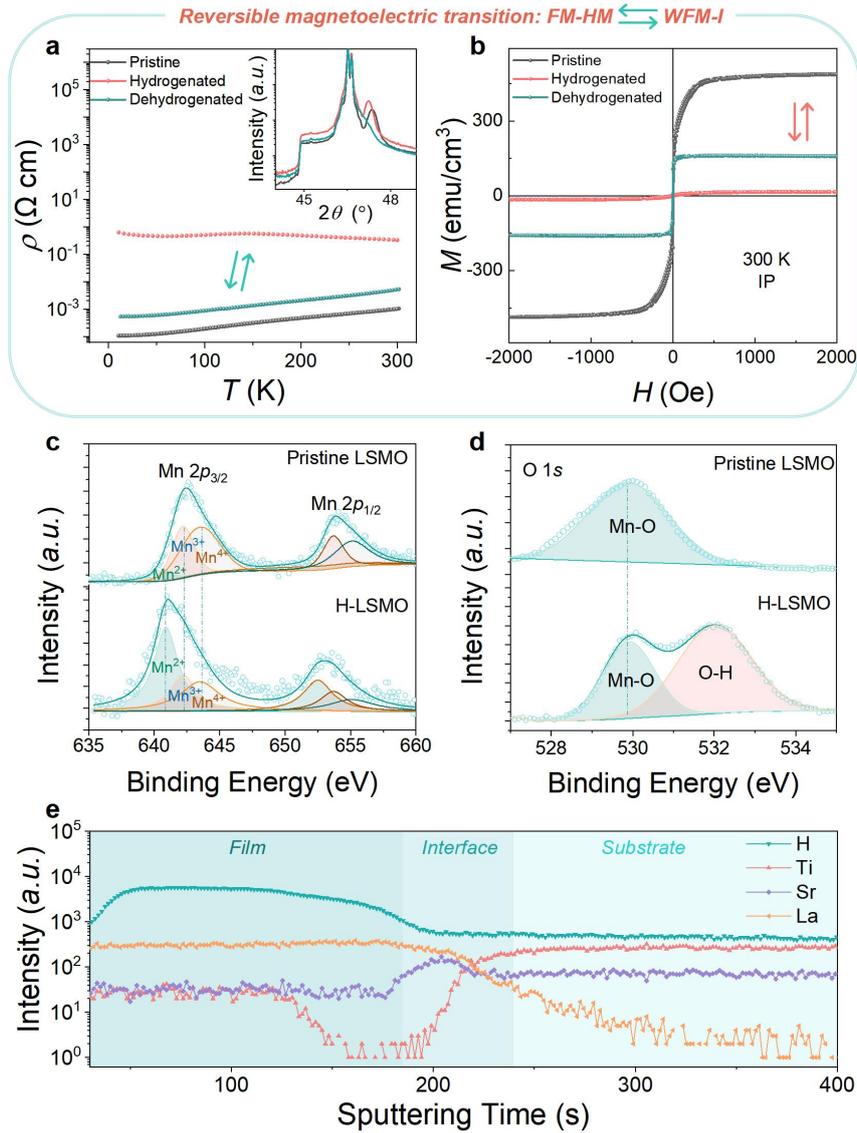

**Figure 3. Ion-charge-spin coupling in hydrogenated LSMO. a**, *M-T* tendency compared for LSMO film under hydrogenation and dehydrogenation. **b**, Comparing the *M-H* loops for hydrogenated and dehydrogenated LSMO. **c-d**, X-ray photoelectron spectra (XPS) compared for the core levels of **c**, manganese and **d**, oxygen for LSMO through hydrogenation. **e**, Depth profile of the elementary concentration for hydrogenated LSMO/STO (001) heterostructure using time-of-flight secondary ion mass spectrometry (ToF-SIMS) technique.

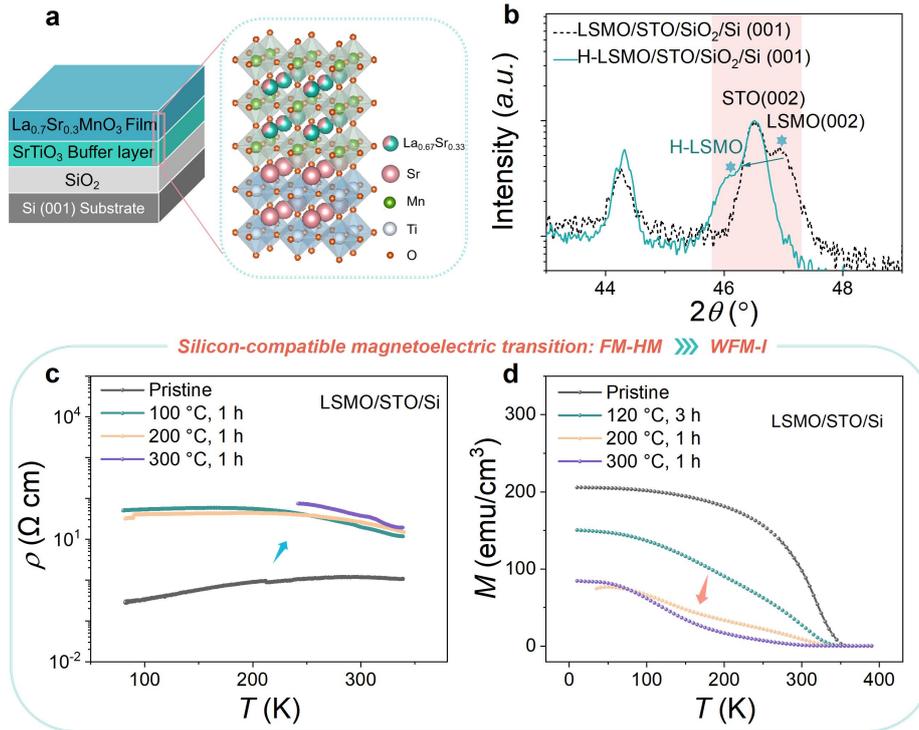

**Figure 4. Silicon-compatible hydrogenated magnetoelectric states in LSMO. a**, Schematic of heterogeneous integration for hydrogenated LSMO on silicon using STO epitaxial buffer layer. **b**, Comparing the XRD pattern for LSMO integrated on silicon before and after hydrogenation at 300 ºC for 1 h. **c**, Electronic state evolution for LSMO/STO/Si heterostructure via hydrogenation. **d**, Hydrogen-triggered magnetic phase modulations in LSMO film integrated on silicon.

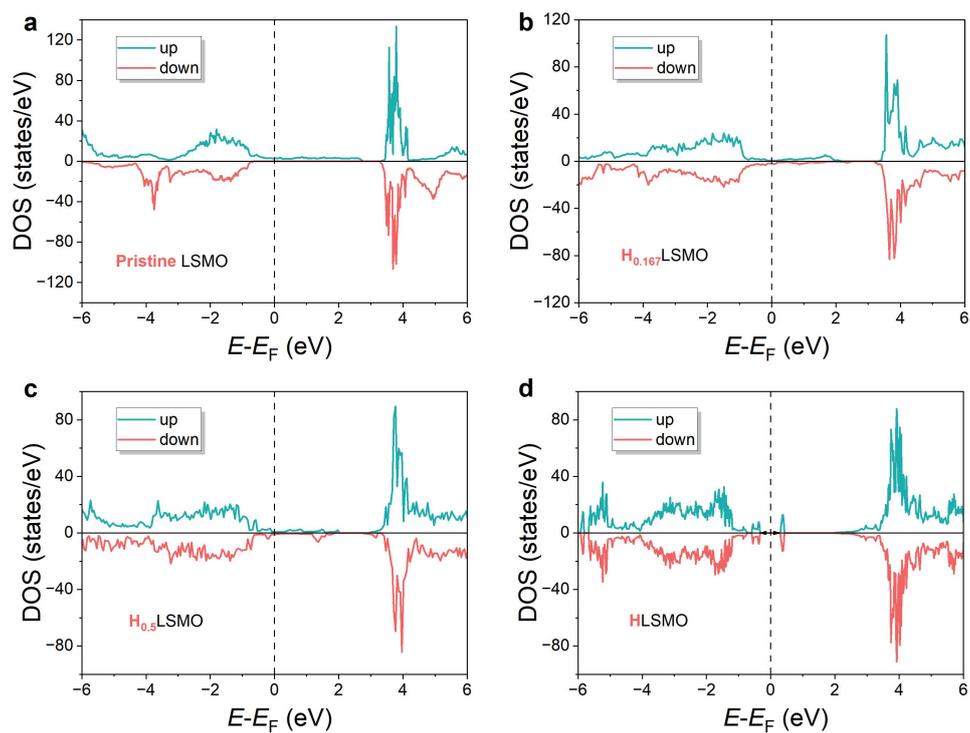

**Figure 5. Calculating the electronic band structure for LSMO system through hydrogenation.** Calculated density of states (DOS) of LSMO system **a**, pristine, **b**, $H_{0.167}$LSMO, **c**, $H_{0.5}$LSMO and **d**, HLSMO.